\documentstyle[aps,prl]{revtex}
\begin{document}
\twocolumn[\hsize\textwidth\columnwidth\hsize\csname
@twocolumnfalse\endcsname

\draft
\title{Quantum information processing in localized modes of light within a photonic
band-gap material}

\author{Nipun Vats\cite{emailadd}, Terry Rudolph and Sajeev John}
\address{Department of Physics, University of Toronto, 60 St. George Street, Toronto,
Ontario, M5S 1A7, Canada}
\date{October 11, 1999}
\maketitle
\widetext
\begin{abstract}
The single photon occupation of a localized field mode within an engineered network
of defects in a photonic band-gap (PBG) material is proposed as a unit
of quantum 
information (qubit). Qubit operations are mediated by optically-excited atoms
interacting with these localized states of light as the atoms traverse the
connected void network of the PBG structure. We describe conditions
under which this system can have independent qubits with controllable
interactions
and very low decoherence, as required for quantum
computation.
\end{abstract}
\pacs{PACS numbers: 03.67.-a, 42.70.Qs, 42.50.Ct }

\vfill
\narrowtext

\vskip2pc]

The field of quantum information has experienced an explosion of interest
due in large part to the creation of quantum algorithms that are far more
efficient at solving certain computational problems than their classical
counterparts \cite{qcreview}. However, despite the development of error
correction protocols designed to preserve a desired quantum state
\cite{knill} and several promising proposals for the implementation of
quantum information processing \cite{qcproposals}, experimental progress
to date has been limited to systems of only 2 or 3 quantum bits (qubits).
This is due to the difficulties associated with the precise preparation
and manipulation of a quantum state, as well as decoherence; that is, the
degradation of a quantum state brought about by its inevitable coupling to
the degrees of freedom available in its environment.

In this Letter, we propose a scheme for quantum information processing in
which our qubit is the single photon occupation of a localized defect mode
in a three--dimensional photonic crystal exhibiting a full photonic
band-gap (PBG). The occupation and entanglement of multiple qubits is
mediated by the interaction between these localized states 
and an atom with a radiative transition that is nearly resonant with the
localized modes. The atom passes between defects through a
matter--waveguide channel in the extensive void network of the PBG
material. We argue that such materials may provide independent qubit
states, low error qubit (quantum gate) operations, a long decoherence time
relative to the time for a gate operation, the potential for scalability
to a large number of qubits, and considerable flexibility in the means of
controlling the quantum state of the system.

PBG materials have themselves been an area of intensive research over the past
decade \cite{crete}. These are highly porous, three--dimensionally periodic
materials of high refractive index with pore periodicity on the length scale
of the relevant wavelength of light. The combination of Bragg scattering from
the dielectric backbone and Mie scattering resonances of individual voids leads
to the complete exclusion of electromagnetic modes over a continuous range of
frequencies. A PBG material amenable to fabrication at microwave and optical
wavelengths is the stacked wafer or ``woodpile''
structure, constructed in analogy with the arrangement of atoms in crystalline
silicon \cite{woodpile}. An optical PBG material is the inverse
opal structure \cite{invopal},
which is a self--organizing FCC array of connected air spheres in a high index
material such as Si, Ge, or GaP (Fig. 1).  The void
regions in both of these structures allow
line of sight propagation of an atomic beam through the crystal. For example,
the Si inverse opal consists of approximately \( 75\% \) connected
void regions,
can be readily grown on the scale of hundreds of lattice constants, and can
exhibit a PBG spanning \( \sim 10\% \) of the gap center frequency. 

An atom with a transition in a PBG will be unable to spontaneously emit a
photon; instead, a long--lived photon--atom bound state is formed
\cite{john&wang}. By introducing isolated voids that are larger (air
defect) or smaller (dielectric defect) than the rest of the array,
strongly localized single mode states of light can be engineered within
the otherwise optically empty PBG
\cite{defectrefs,joannopoulos,lin_defect}. For isolated defect modes in a
high quality dielectric material, Q-factors of \( 10^{10} \) or higher
should be attainable (see below). Excited atoms passing through or near a
defect can exchange energy coherently with this localized state, thus
preparing the state of our qubit. Furthermore, such ``point defects'' in
PBG materials can give rise to modal confinement to within the wavelength
of the mode \cite{lin_defect}, giving an enhancement of the cavity Rabi
frequency of \( 5-50 \) times over conventional microcavities. These facts
make localized modes in PBG materials excellent candidates for the strong
coupling regime of cavity quantum electrodynamics (CQED)
\cite{haroche,kimble}.

Along with the point defects described above, line defects (waveguides)  
can be engineered within a PBG material. This can be accomplished by
modifying the initial templating mold of a single inverted opal crystal
\cite{templating}, or by removing or modifying selected dielectric rods in
a woodpile crystal. Such extended defects can be used to inject a coherent
light field into the PBG of the system, thereby controllably altering the
Bloch vector of a two--level atom that passes through the illuminated
region without the generation of unwanted quantum correlations in the
system. The information thereby input to the system can then
be transferred from the atom to the localized modes. Our qubits are
protected from the narrowly confined externally injected fields by the
surrounding dielectric lattice.

Our basic scheme for the entanglement of localized modes is shown in Fig.
1. A two--level atom of excited state \( \left| e\right\rangle \) and
ground state \( \left| g\right\rangle \) is prepared in an initial state
\( \left| \psi \right\rangle =c_{e}\left| e\right\rangle +c_{g}\left|
g\right\rangle \) in an illuminated line defect as it passes through the crystal with velocity \( v \)
(path A).  It then successively interacts with localized states $p$, $q$,
and $p'$. The state of the atom may be further modified in mid-flight as
it passes though additional line defects. Due to the absence of
spontaneous emission in a PBG, the coherence of the atom mediating the
entanglement may be maintained over many defect spacings. This allows for
the entanglement of more than two defects or of pairs of more distant
defects with lower gate error than is possible with conventional
microcavity arrays. The lack of spontaneous emission may also enable the
use of atomic or molecular excitations that would be too short--lived for
use in conventional CQED.  Measurements of the states of defect modes can
be made via ionization measurements of the state of a probe atom after it
has interacted with individual defects (e.g. path B).

In principle, our proposal applies to systems with atomic transitions
in the microwave or in the optical/near-IR.  However, to
perform a precise sequence of many gate operations will require that single
atoms of known velocities be sent through the crystal with known
trajectories at well--defined times.  In a microwave PBG material, the
void channels are sufficiently large ($\sim 3{\rm mm}$) that
experiments involving small numbers of qubits
can be performed with atoms velocity selected from a thermal source which
initially emits atoms with a Poissonian velocity distribution
\cite{haroche}.  The outcome of these simple quantum algorithms can then be
evaluated by statistical measures on an ensemble of appropriately prepared
atoms passing through the defect network.  Because optical PBG materials
have micron--sized void channels, atomic waveguiding may be necessary in
order to prevent the van der Waals adhesion of the atoms to the dielectric
surfaces of the crystal.  To this end, a field mode excited from the lower
photonic band edge resides almost exclusively in the dielectric fraction
of the crystal, producing an evanescent field at the void--dielectric
interface.  This field acts as a repulsive atomic potential if
blue--shifted from an atomic transition \cite{tannoudji}. Similarly, one
may guide atoms with such evanescent waves through engineered line defects
which support only field modes far detuned from the qubit atomic
transition, in a manner analogous to atomic waveguiding in hollow optical
fibers \cite{atomwaveguide}. Using such techniques, optical CQED and atom
interferometry experiments using thermally excited atoms or atoms dropped
into a PBG material from a magneto--optical trap should currently be
possible.  We note that long--lived localized modes may also be used to
entangle the electronic states of ions in ion traps \cite{ion} .  Unlike
with atomic beams, trapped ions may be precisely
manipulated into and out of a defect mode in a PBG
material.  This may make the ionic system more amenable to large scale
implementations of quantum information.

We first consider the dynamics of a two--level atom passing through a point
defect. In the dipole and rotating wave approximations, the evolution of an
atom at position \( {\mathbf r} \) in a localized defect mode is given by a
position dependent Jaynes--Cummings Hamiltonian \cite{kurizki,buzek},
\begin{equation}
\label{jaynes}
H({\mathbf r})=\frac{\hbar \omega _{a}}{2}\sigma _{z}+\hbar \omega
_{d}a^{\dagger }a+\hbar G({\mathbf r})\left( a\sigma _{+}+a^{\dagger }\sigma _{-}\right) ,
\end{equation}
 where \( \sigma _{z} \), \( \sigma _{\pm }=\sigma _{x}\pm i\sigma _{y} \)
are the usual Pauli spin matrices for a two--level atom with transition frequency
\( \omega _{a} \), and \( a \), \( a^{\dagger } \) are respectively the annihilation
and creation operators for a photon in a defect mode of frequency \( \omega _{d} \).
The atom--field coupling strength may be expressed as $ G({\mathbf
r})=\Omega _{0}\left( \hat{{\mathbf d}}_{21}\cdot \hat{{\mathbf
e}}({\mathbf r})\right) f({\mathbf r}) $ ,
where \( \Omega _{0} \) is the peak atomic Rabi frequency over the defect
mode, \( \hat{{\mathbf d}}_{21} \) is the orientation of the atomic dipole moment,
and \( \hat{{\mathbf e}}({\mathbf r}) \) is the direction of the electric field
vector at the position of the atom. In general, the three-dimensional
mode structure will
be a complicated function of the size and shape of the defect
\cite{joannopoulos}.
However, we need only consider the one--dimensional mode profile that
intersects
the atom's linear path, i.e. \( G({\mathbf r})\rightarrow G(r) \). The profile
\( f(r) \) will have an exponential envelope centered about the point in the
atom's trajectory that is nearest to the center of the defect mode, \( r_{0} \).
Within this envelope, the field intensity will oscillate sinusoidally, and for
fixed dipole orientation, variations in the relative orientation of the dipole
and the electric field will also give a sinusoidal contribution \cite{comment_dipole}.
For a PBG material with lattice constant \( a \), we can thus set \( \hat{{\mathbf d}}_{21}\cdot \hat{{\mathbf e}}(r)=1 \)
and write $f(r)=e^{-\left| r-r_{0}\right| /R_{\rm def}}\cos
\left[ \frac{\pi }{a}(r-r_{0})+\phi \right] $ (see Fig. 2).
Because we want to transfer energy between the atom and the localized mode,
we wish to use modes that are highly symmetric about the atom's path.
We therefore set 
\( \phi =0 \). \( R_{\rm def} \) defines the spatial extent of the mode, and is
at most a few lattice constants for a strongly confined mode deep in a
PBG \cite{joannopoulos}.

The atom--field state function after an initially excited atom has passed through
a defect can be written in the form
\begin{equation}
\label{finalstate}
\left| \Psi _{f}\right\rangle =u\left| e,0\right\rangle +w\left| g,1\right\rangle .
\end{equation}
 \( u \) and \( w \) are obtained by replacing \( r-r_{0} \) by \( vt-b \)
in the Hamiltonian (\ref{jaynes}) and integrating the corresponding
Schr\"odinger
equation from \( t=0 \) to \( 2b/v \).
\( b \) is chosen so that the interaction is negligible at \( t=0 \); we set
\( b=10R_{\rm def} \). Fig. 2 plots \( \left| u\right| ^{2} \) for the
$\lambda=780{\rm nm}$
transition of an initially excited Rb atom traveling through
a defect in an optical PBG material (e.g., GaP) at
thermally--accessible velocities ($v\sim 100 - 600 {\rm m/s}$)
for various detunings of the atomic
transition frequency from a defect resonance, \( \delta \equiv \omega
_{a}-\omega _{d} \).
Such a detuning can be achieved by applying an external field to Stark shift
the atomic transition as the atom passes through selected defects. 
The final
state of the atom is seen to be a sensitive function of \( \delta
\) and $v$.  Similar
velocity--dependent atomic inversions are obtained for the above
system with the atom in free fall ($v\sim .1{\rm m/s}$), and for the
$5.9{\rm mm}$ Rydberg transition of Rb using a thermal beam of atoms passing
through an appropriate microwave PBG material.  At
thermal velocities, the
inversion can be finely tuned within the $1{\rm m/s}$ resolution of thermally
generated atomic beams.

To show that our system is capable of encoding quantum algorithms for realistic
values of system parameters, we demonstrate the viability of producing a maximally
entangled state of two defect modes after an atom prepared in its excited state
has passed through both defects. The final state of this system after the two
defect entanglement can be written as
\begin{equation}
\label{state_entangle}
\left| \Psi \right\rangle =\alpha \left| g,1,0\right\rangle +\beta \left| g,0,1\right\rangle +\gamma \left| e,0,0\right\rangle ,
\end{equation}
 where \( \left| i,j,k\right\rangle  \) refers to the state of the atom and
the photon occupation number of the first and second defect modes respectively.
\( \alpha  \), \( \beta  \) and \( \gamma  \) can be expressed in terms of
the probability amplitudes for an atom that has passed through only defect 1
or 2; in the notation of Eq. (\ref{finalstate}), \( \alpha =w(1) \), \( \beta =u(1)w(2) \),
and \( \gamma =u(1)u(2) \). Maximal defect entanglement is obtained for \( \left| \alpha \right| =\left| \beta \right| =1/\sqrt{2} \),
\( \left| \gamma \right| =0 \), which leaves the atom disentangled from the
defect modes. As an example, using the system of Fig. 2 with $v=278{\rm m/s}$
and with the atom on
resonance with the first defect, maximally entangled states are
obtained for
atomic detunings of 
\( \delta _{2}/\Omega _{0}=.07 \), \( .13 \) and \( .31 \) from the
second defect mode.

As discussed, by passing a 2--level atom through an externally
illuminated line defect, the atom's Bloch vector can be
initialized or modified in--flight without becoming correlated with
the state of a defect mode \cite{careful}.  For simplicity, we assume that an atom
sees a uniform mode profile as it crosses such an optical
waveguide mode of
width $2\lambda$.  For an injected single--mode field resonant with
the atomic transition, the
Bloch vector will then rotate at the semi--classical Rabi frequency,
\( {\cal R} _{0}\equiv {\mathbf d}_{21}\cdot {\mathbf E}/2\hbar  \),
where \( {\mathbf E} \)
is the amplitude of the applied electric field \cite{meystre}. At a thermal
velocity of $100{\rm m/s} $, the minimum field strength, $E$,
required to fully
rotate the
Bloch vector for the $\lambda = 5.9{\rm mm}$ transition in Rb
($ d_{21}=2.0 \times 10^{-26}{\rm C \cdot m} $)  is $ E \sim 1{\rm mV/m} $.  For  
the $\lambda = 780{\rm nm} $ optical transition of Rb ($ d_{21}= 1.0 \times
10^{-29}{\rm C \cdot m} $), $E \sim 9{\rm kV/m} $ at
$v=100{\rm m/s}$, whereas in free fall ($v=.3{\rm m/s}$),
$E\sim 30 {\rm V/m}$.  The weak RF field required in the microwave system
implies that such experiments must be conducted at low temperatures
($\lesssim .6{\rm K}$) in order to prevent the modification of the
atomic state by thermal photons.  In the optical, the required field
strengths are attainable using a cw laser whose output is coupled into
waveguide channels of $\sim \lambda^2$ cross section, and are well below the
ionization field strengths of both the messenger atoms and the PBG material.  

For quantum computation, a universal 2 qubit gate can be constructed in analogy
with conventional CQED \cite{sleator,qcreview}. We outline the construction
of the associated controlled-NOT (CNOT) gate, which flips the occupation state of a
target defect \( p \) only if a second control defect \( q \) is occupied
(Fig. 1b). The state of defect \( p \) is first transferred to an incident atom
via a near--resonant atom--defect interaction, such that \( (\alpha \left| 1\right\rangle _{p}+\beta \left| 0\right\rangle _{p})\left| g\right\rangle \rightarrow (-i\alpha \left| e\right\rangle +\beta \left| g\right\rangle )\left| 0\right\rangle _{p} \).
This is analogous to an integrated \( \pi /2 \) interaction in a spatially
uniform cavity, i.e., one for which \( G({\mathbf r})\rightarrow
\Omega _{0} \). The atomic Bloch vector is then rotated by \( {\cal
R}_0t= \pi /4 \)
by applying a classical field in line defect \( R_{1} \), thus generating
the transformation \( {\cal H}\left| e\right\rangle =(\left| e\right\rangle +i\left| g\right\rangle )/\sqrt{2} \),
\( {\cal H}\left| g\right\rangle =(i\left| e\right\rangle +\left| g\right\rangle )/\sqrt{2} \).
Next, a dispersive interaction \cite{nonres} is created by detuning the
atom far from the defect resonance (but still within the PBG) as it passes through
\( q \). This is used to produce a phase rotation of \( \pi  \) on the excited
state, conditioned on the presence of a photon in \( q \), thus causing the
amplitude of \( \left| e\right\rangle  \) to change sign if \( q \) is occupied.
The Bloch vector rotation \( {\cal H}^{-1} \) is then performed in \( R_{2} \),
which switches states \( \left| e\right\rangle  \) and \( \left| g\right\rangle  \)
relative to their initial values only if a photon was present in \( q \). Finally,
the state of the atom is transferred to defect \( p' \) by a near--resonant
\( 3\pi /2 \) interaction, leaving the atom in its ground state. \( p' \)
then carries the result of the CNOT operation. 

We now turn to the crucial issue of the decoherence and energy loss of a photon
in a defect mode. In general, these processes can have very different time scales,
as the former is a result of the entanglement of a quantum state with its environment,
which can occur well before the dissipation of energy \cite{decohere}. It has
however been shown that for the (linear) coupling between a photon and a non--absorbing
linear dielectric, the coherence of a small number of photons in a given mode
is not destroyed by the interaction \cite{zurek}. Therefore, the coherence
of excited defect modes in a high quality PBG material is essentially limited by energy loss.  We
note that phonon
mediated spontaneous Raman and Brillouin scattering of photons out of a defect
mode are ineffective loss mechanisms due to the vanishing overlap between a
localized field mode and the extended states of the electromagnetic continuum
\cite{john&wang}. Unlike most qubit elements, photons do not interact
significantly with
one another, preventing the propagation of single bit errors through a network
of defect modes. However, a photon may ``hop'' from one defect to another
either via direct tunneling or through phonon--assisted hopping \cite{crete}.
The likelihood of both of these processes decreases exponentially with the spatial
separation of the defects, and is negligible for a defect separation of \( \sim 10 \)
lattice constants for a strongly confined mode. 

For an isolated point defect engineered well inside a large--scale PBG
material, the Q--factor
will then be limited by absorption and impurity scattering from the
dielectric backbone
of the crystal \cite{sakoda}. 
In the optical and microwave frequency regimes, away from the electronic gap
and the restrahlen absorption frequencies of a high quality semiconductor material,
\( \epsilon _{2} \) can be as low as \( 10^{-9} \) at room temperature, and
may be reduced at lower temperatures \cite{palik}. One can further
reduce absorptive losses by minimizing the fraction
of the mode in the dielectric (e.g., a strongly localized mode in an
air defect).
Judicious defect fabrication in a high quality dielectric material should then
give Q--factors of \( 10^{10} \) (assuming 10\% of the mode is in the
dielectric)
or higher, which corresponds to photon lifetimes of \( 10^{-1}{\rm sec} \) and \( 10^{-4}{\rm sec} \)
at microwave and optical frequencies respectively. This Q--value is comparable
with present microwave CQED experiments, and is in excess of that currently
used in optical CQED.  Assuming defects are separated
by 10 lattice constants, we obtain a
decoherence to gate time ratio of $\sim 200$ even at room temperature,
showing that our system
is potentially capable of encoding complex quantum algorithms.

We thank K. Busch, H.M. van Driel, and J.M. Raimond for valuable
discussions.  N.V. acknowledges support from the Ontario Graduate
Scholarship Program.  This work was sponsored in part by the 
Killam Foundation, 
the New Energy and
Industrial Technology Organization (NEDO) of Japan, Photonics Research
Ontario, and the NSERC
of Canada.

\begin{figure}[h]
\caption{(a) Schematic diagram of atomic trajectory (dashed line)
through a void channel in an inverted opal PBG material.  A localized
air defect is made by enlarging one of the spherical voids along the
atom's path.  (b)  Defect arrangement for a CNOT operation.  Local
defects $p$, $q$, and $p^{\prime}$ are separated by line defects ${\rm
R_1}$ and ${\rm R_2}$, through which coherent fields are applied as
the atom moves along path A.  
}
\label{fig1}
\end{figure}

\begin{figure}[tbp]
\caption{Excited atomic state probability, $\left| u \right|^2$, after an
initially excited Rb atom ($\omega_a=2.4 \times 10^{15}{\rm rad/s}$) has interacted
with a point defect of $R_{\rm def}=a$,
$\phi=0$; $a=0.8(2\pi c/\omega_a)$. We take $\Omega_0 = 1.1\times 10^{10}{\rm rad/sec}$
, which assumes the mode is confined within a single wavelength.
$\delta=\omega_a - \omega_d$.  Inset:
corresponding mode profile, $f(r)$.}
\label{fig2}
\end{figure}


\begin{thebibliography}{}
\bibitem{comment_dipole} Rotation of $\hat{\mathbf d}_{21}$ will result in a
modification of the atom--defect interaction without the spontaneous
decay expected in a planar microcavity.
\bibitem{careful} Line defects must be engineered such that 
they do not enhance atomic spontaneous emission.
\end{thebibliography}
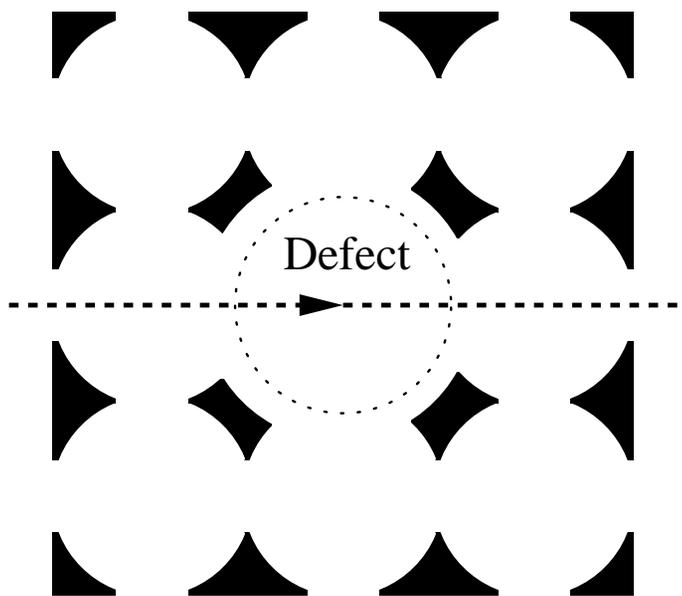
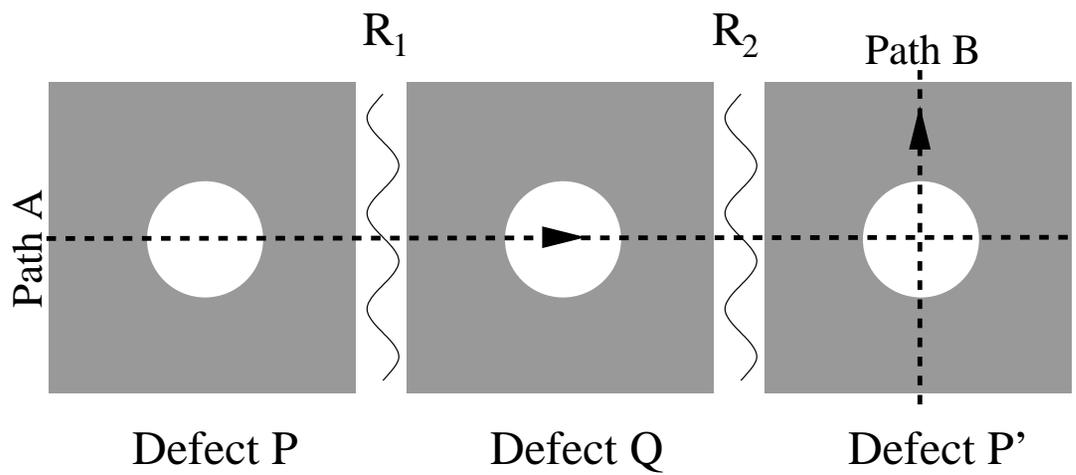
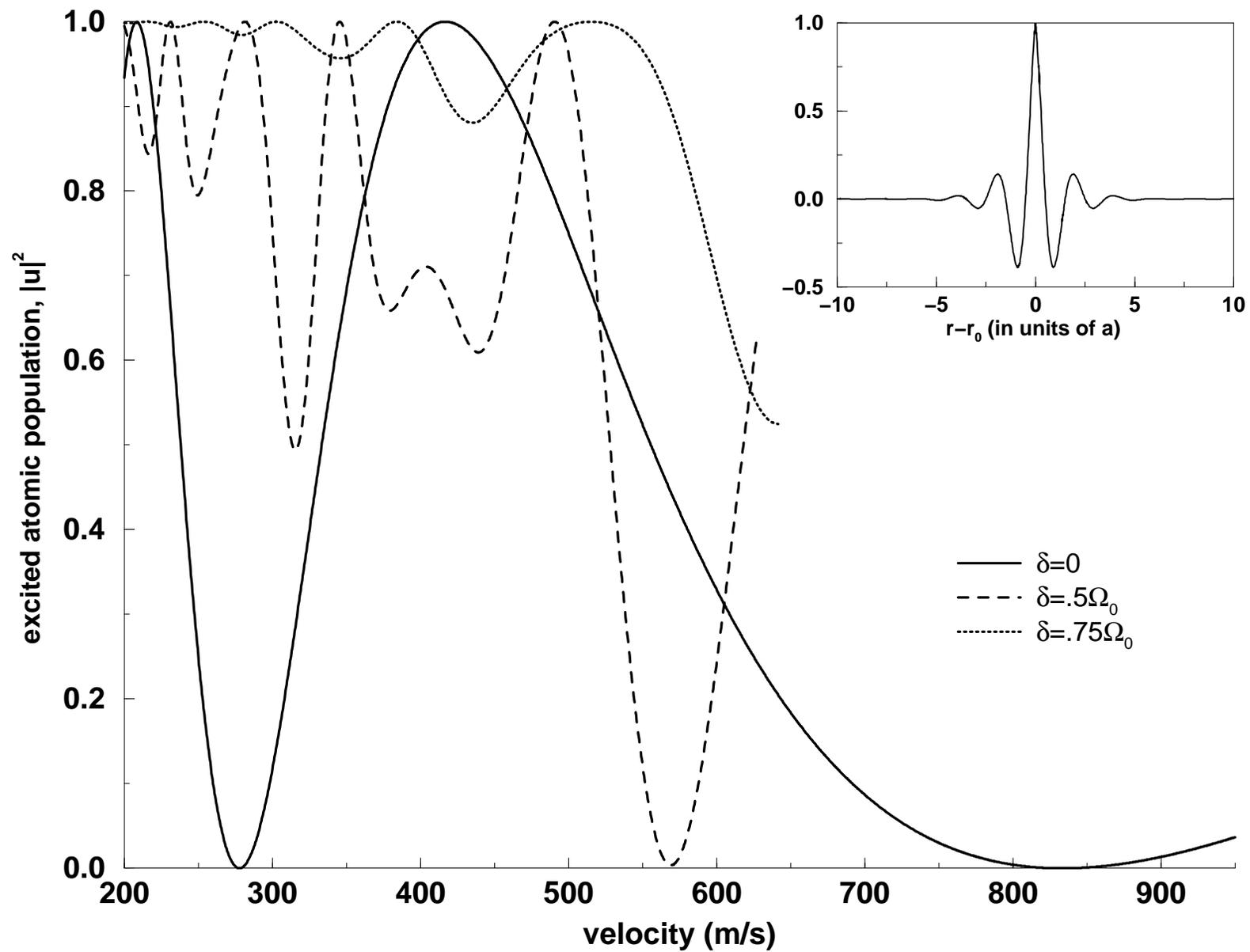
\end{document}